\documentclass{moriond}

\def\be{\begin{equation}}
\def\ee{\end{equation}}
\def\bea{\begin{eqnarray}}
\def\eea{\end{eqnarray}}

\newcommand{\Photo}{}

\begin{document}
\vspace*{4cm}
\title{Resummation Scales and the Assessment of 
Theoretical Uncertainties in \\  
Parton Distribution Functions~\footnote{Presented  
by F.~Hautmann at the {\it 29th Workshop on Deep-Inelastic Scattering 
and Related Subjects}, 2-6 May 2022.}}

\author{V.~Bertone$^1$,   G.~Bozzi$^{2,3}$ and F.~Hautmann$^{4,5,6}$}

\address{$^1$IRFU, CEA, Universit\'e Paris-Saclay, F-91191 Gif-sur-Yvette\\
$^2$Dipartimento di Fisica, Universit\`a di Cagliari, Cittadella Universitaria, I-09042 Monserrato\\  
$^3$INFN, Sezione di Cagliari, Cittadella Universitaria, I-09042 Monserrato\\  
$^4$CERN, Theoretical Physics Department, CH 1211 Geneva \\ 
$^5$Elementaire Deeltjes Fysica, Universiteit Antwerpen, B 2020 Antwerpen\\ 
$^6$Theoretical Physics  Department, University of Oxford, 
Oxford OX1 3PU
}

\maketitle

\abstracts{We discuss perturbative solutions  of renormalization group equations, and  
propose the use of resummation scale techniques in assessing 
 theoretical uncertainties  on the  extraction   of  parton distribution functions from data.}

\vspace*{-8cm} 
\begin{flushright} 
CERN-TH-2022-088 
\end{flushright} 
\vspace*{8cm}

Many applications of Quantum Chromodynamics (QCD) to precision calculations for 
high-energy collider physics involve the perturbative solution of renormalization group equations (RGEs). 
It has long been known that, at any given order of 
perturbation theory,  the differences  between RGE solutions  
obtained by methods which 
are equivalent up to 
formally subleading orders 
provide a significant source of theoretical uncertainty on predictions for 
physical observables. 

In the literature of calculations based on  QCD resummation, 
these uncertainties are commonly taken into account by a variety 
of different techniques, which appeal to the principle of introducing 
so-called resummation scales, of the order of the hard momentum-transfer 
scale of the process but otherwise arbitrary, and setting criteria to let them 
vary and evaluate the corresponding variation 
in  
the theoretical prediction. 

In several other common applications of QCD to collider physics,  on the 
other hand,  
resummation scale variations are not performed. 
A notable example is the determination of 
parton distribution functions (PDFs)  from global fits to experimental 
data,~\cite{Ball:2022hsh,Ball:2021leu,NNPDF:2019ubu,Bailey:2020ooq,Hou:2019efy,Alekhin:2017kpj,Abramowicz:2015mha} 
central to virtually any aspect of hadron collider physics.  
In this application, RGEs and their perturbative solutions are used 
for the evolution of the PDFs themselves 
and for the evolution of the running coupling.  
The fact that RGE theory uncertainties are not 
included and resummation scale variations are not performed leads to 
an underestimate of the uncertainties associated with the 
extraction of PDFs from global fits to data. 

This point is illustrated in detail in our recent work~\cite{Bertone:2022sso}.  
There, we discuss various aspects of theory uncertainties  
from RGE perturbative solutions   
and propose  resummation scale methods to characterize them. Applying this 
to  deep-inelastic 
scattering (DIS) structure functions, we show that resummation scale effects 
are significant  in  kinematic regions  relevant to PDF determinations.   
In this article we summarize some of the results from Ref.~\cite{Bertone:2022sso}. 

We consider RGE of the general form 
\begin{equation} 
\label{eq:RGEproto}
  \frac{d\ln R}{d\ln \mu} (\mu, \alpha_s(\mu) )= \gamma(\alpha_s(\mu))\,,
\end{equation}
where $R$ is a renormalized quantity, function of the strong coupling 
$\alpha_s$ and renormalization scale $\mu$, and  $\gamma$ is 
its anomalous dimension, computable as a 
power series expansion in the coupling  $\alpha_s$. 

In Ref.~\cite{Bertone:2022sso} we examine several examples,  
at next-to-leading  (NLO) and next-to-next-to-leading  (NNLO) order,  
and analyze  
RGE solutions obtained by methods which differ by subleading-order 
terms.\footnote{The examples considered in 
Ref.~\cite{Bertone:2022sso} are single-logarithmic resummation problems. 
Similar effects in double-logarithmic Sudakov problems have been observed 
e.g.~in Refs.~\cite{Ebert:2021aoo,Billis:2019evv,Hautmann:2019biw}.} 
 To characterize the differences between such solutions, 
we 
 introduce the evolution operator $G$ which connects 
$R$ at two  scales $\mu$ and $\mu_0$, $R_{\mu} = G ( \mu , \mu_0) R_{\mu_0}$, and 
trace the differences between solutions back to the fact that, due 
to subleading orders, one has, for  a given $\mu^\prime$,    
$G (\mu, \mu_0) \neq G (\mu, \mu^\prime) G (\mu^\prime , \mu_0)$.  
We refer to this as {\it perturbative hysteresis}~\cite{Bertone:2022sso}.   

For the application to PDF evolution, $R$ in 
  Eq.~(\ref{eq:RGEproto}) is to be thought of as the 
  flavor multiplet of PDF Mellin moments, 
and   $\gamma$ as the Mellin transform of 
DGLAP splitting functions. Besides, one has an RGE of the 
form (\ref{eq:RGEproto}) for the running coupling, in which  
 $R$ is proportional to the 
coupling $\alpha_s$, and $\gamma$ to the QCD $\beta$ 
function. Hysteresis effects in both these RGEs 
are analyzed quantitatively in Ref.~\cite{Bertone:2022sso}. 
 
The extraction of  PDFs  $f_j ( x , \mu)$ from collider data 
 uses  factorization formulas relating $f_j$ to physical observables $\Sigma$, 
 e.g.~DIS structure functions,    with the schematic form 
\begin{equation}
\label{eq:factS} 
\Sigma ( x , Q) = \sum_j \int dz H_j ( z, Q, \alpha_s ( \mu_R) ,  \mu_F) f_j 
( x / z, \mu_F )  , 
\end{equation}   
where $H_j$ are perturbatively  computable hard-scattering functions, 
$\mu_F$ is the factorization scale, and $\mu_R$ is the renormalization 
scale. Standard QCD calculations estimate theoretical uncertainties 
associated with unknown higher orders in Eq.~(\ref{eq:factS}) 
by setting criteria for variations  of the scales $\mu_F$ and $\mu_R$. 
 
Ref.~\cite{Bertone:2022sso}  observes  that theoretical uncertainties 
on the extraction of PDFs arise both from $\mu_F$ and $\mu_R$ 
variations in Eq.~(\ref{eq:factS})  and from unknown higher orders in 
the kernels $\gamma$ in  Eq.~(\ref{eq:RGEproto}), and  
  proposes to take the latter into account 
by resummation scale techniques.  To  this end, two different  
but essentially equivalent methods are formulated. One is based on 
expressing the evolution 
operator $G$ via the  
analytic formalism of 
$g$-functions,~\cite{Bertone:2022sso}  
which is frequently applied in soft-gluon resummation 
calculations.
Schematically, at the $k$-th 
order of logarithmic accuracy ${\rm N}^{k}{\rm LL}$,   
this is of the form 
\begin{equation}
\label{eq:gensoldglap}
 G^{{\rm N}^{k}{\rm LL}} \sim g_0^{ {\rm N}^k{\rm  LL}}(\lambda)\exp\left[\sum_{l=0}^{k} \alpha_s^l(\mu) g_{l+1}(\lambda)\right]  \;\; , \;\;\; 
\lambda \sim \alpha_s(\mu) \beta_{0}\ln (\mu^{\rm (Res)} / \mu )\,,
\end{equation}
where the $g$'s  are perturbatively 
 computable functions of the scaling variable $\lambda$, and  
$\mu_{\rm Res}$ is the resummation scale. 
Explicit expressions for the $g$-functions up to $k=3$ are given 
in Ref.~\cite{Bertone:2022prep}.
 Another method is based on displacing 
 the argument of the coupling appearing in the 
 perturbative expansion of $\gamma$, so as to obtain a 
 new  effective  anomalous dimension  
 differing from the previous one by subleading terms. This method  is well-suited for  
 numerical evaluations.  In either method, one ends up 
 introducing  resummation scales   
 $\mu^{\rm (Res)}_{\rm PDF}$ for the PDF evolution  and 
 $\mu^{\rm (Res)}_{\alpha_s}$ for the coupling evolution.

\begin{figure*}[t!]
\begin{center}
\includegraphics[width=0.43\textwidth]{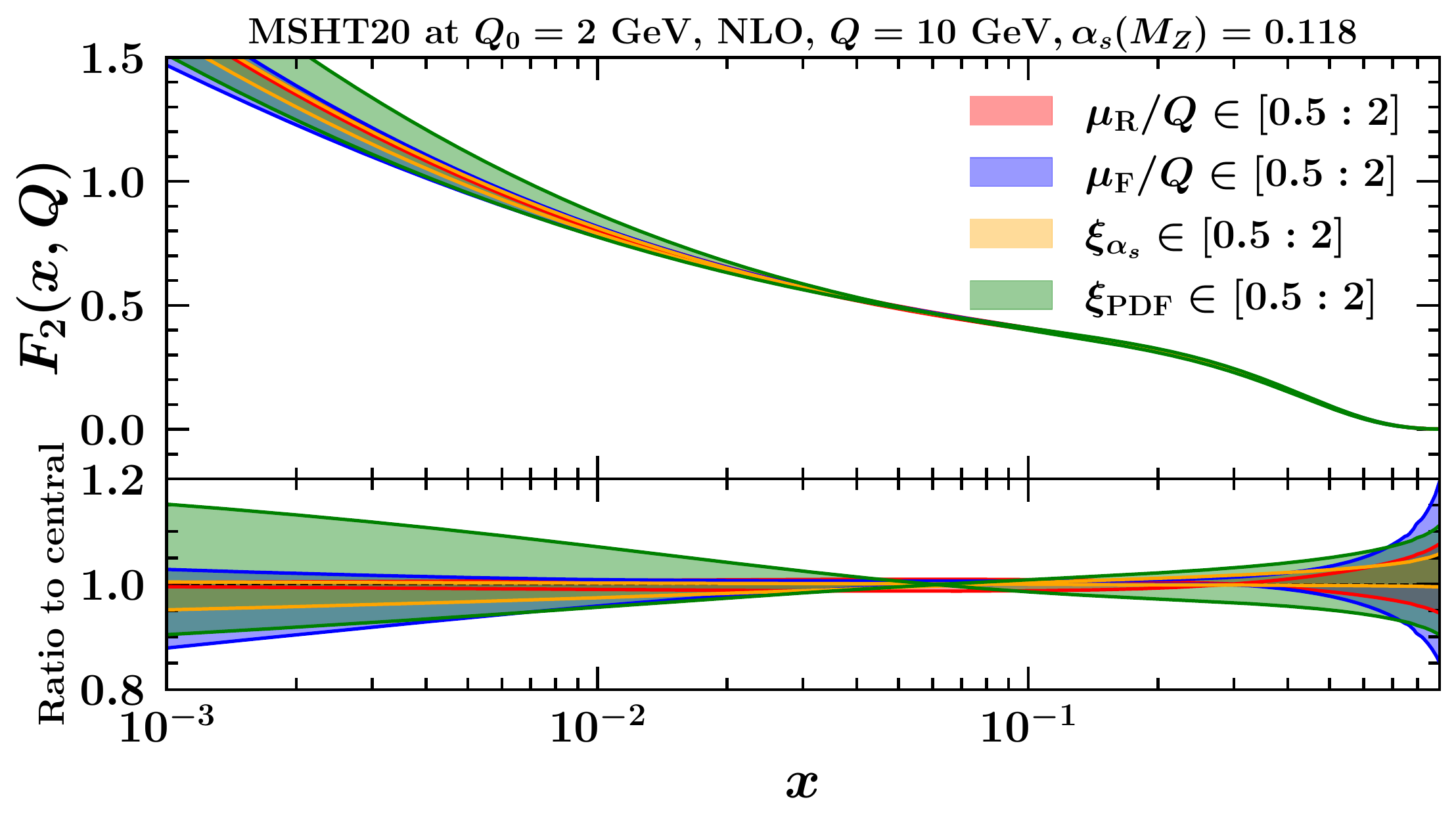}
\includegraphics[width=0.43\textwidth]{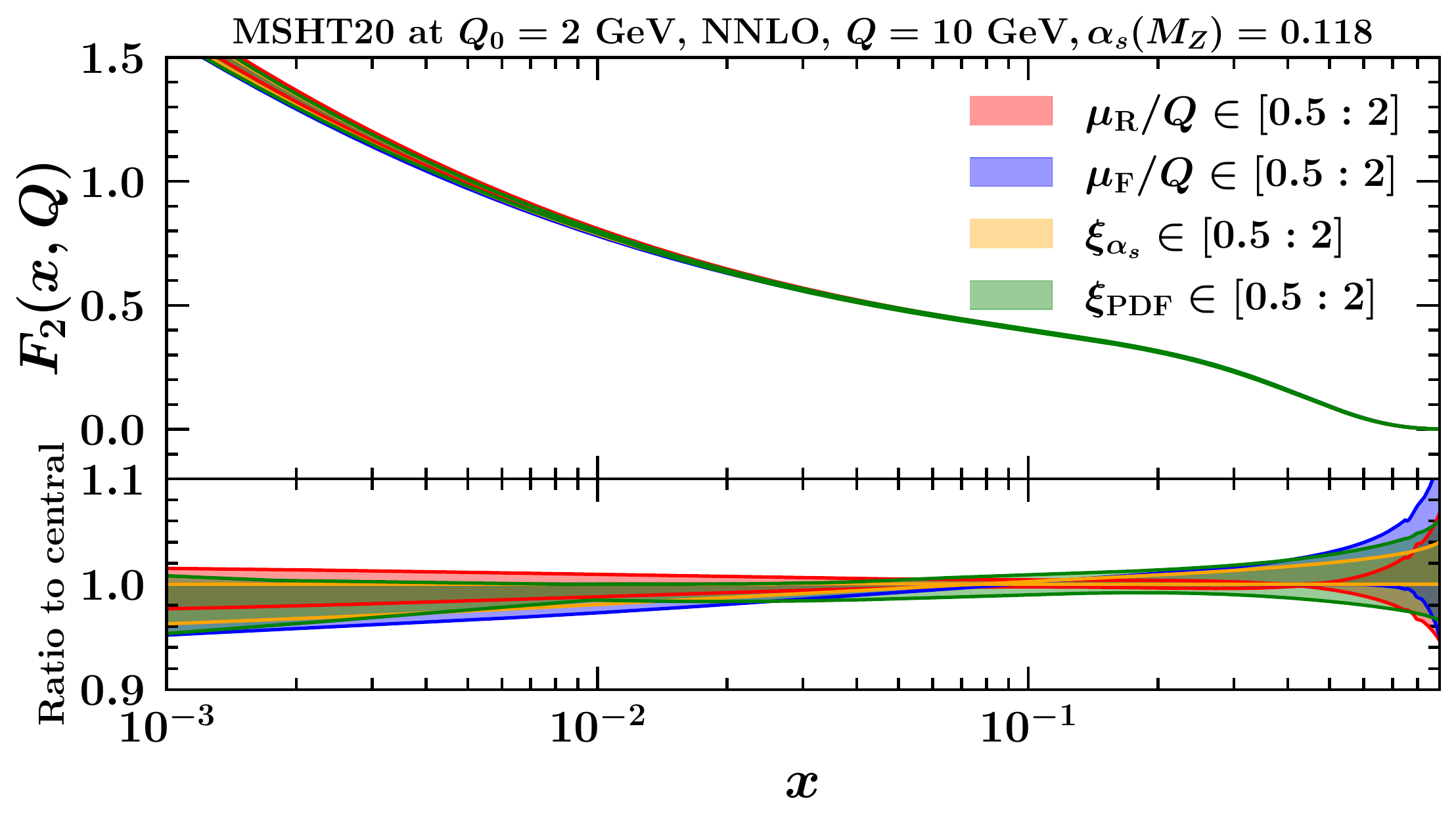}
\caption{The  structure function $F_2$ from Ref.~\protect\cite{Bertone:2022sso} versus $x$, 
at NLO and NNLO in perturbation 
theory, with the uncertainty bands associated with variations of renormalization and 
factorization scales, $\mu_R$ and $ \mu_F$, and resummation scale parameters $\xi_{\alpha_s}$ and $\xi_{\rm PDF}$. 
We use MSHT20 PDFs  at $Q_0 = 2$ GeV and $\alpha_s (M_Z ) = 0.118 $ as RGE inputs.}
\label{fig:f2_muFmuR2resscales1}
\end{center}
\end{figure*}

As an application, we illustrate the case in which  we take $\Sigma$ in Eq.~(\ref{eq:factS})  to be the DIS structure function 
 $F_2 (x , Q)$.   We perform   variations of renormalization and factorization scales, $\mu_R$ and $\mu_F$, and of  
 resummation scales, which we  
  parameterize  as     $\mu^{\rm (Res)}_{\alpha_s} =  \xi_{\alpha_s} Q$ 
 and  $\mu^{\rm (Res)}_{\rm PDF}  =  \xi_{\rm PDF} Q$. 
Fig.~\ref{fig:f2_muFmuR2resscales1}    shows results  at NLO (left panel) and NNLO (right panel). 
Resummation scale 
uncertainties are observed to be generally non-negligible with respect to the 
renormalization and factorization scale uncertainties in the kinematic 
region considered. In 
the left hand side panel,  we see 
that the effect of the 
resummation scale parameter $\xi_{\rm PDF}$ dominates the low-$x$ region while 
the effect of the factorization scale $ \mu_F$ dominates at the largest $x$. 
  The right hand side panel   illustrates  that  
the size of the uncertainty bands is reduced at NNLO. 
The behavior  above    
reflects   
 higher-order corrections to   
$F_2$ at small $x$ being dominated by flavor-singlet 
 quark anomalous dimensions.~\cite{Catani:1993rn,Ellis:1995gv}

\begin{figure*}[t!]
\begin{center}
\includegraphics[width=0.43\textwidth]{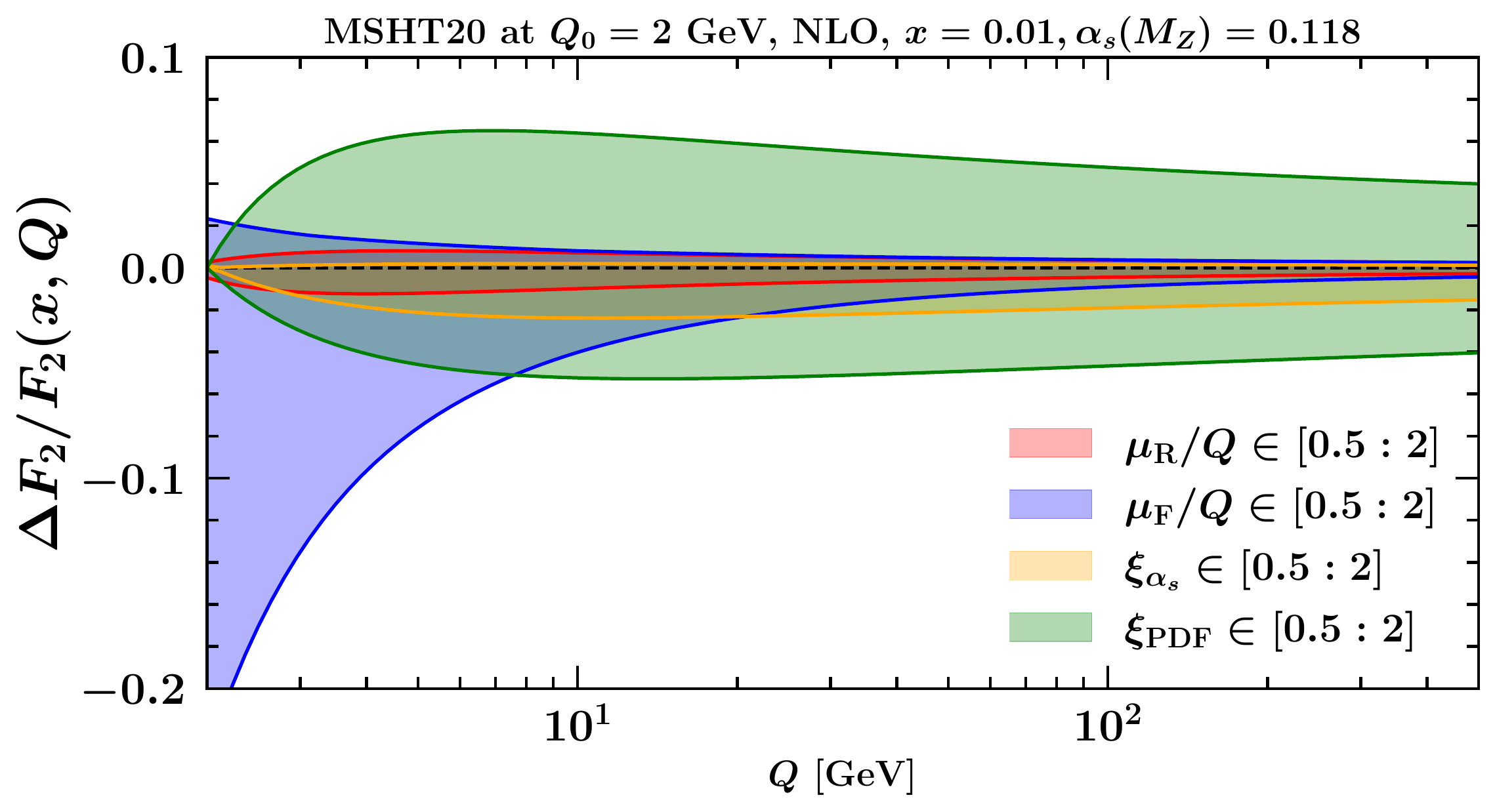}
\includegraphics[width=0.43\textwidth]{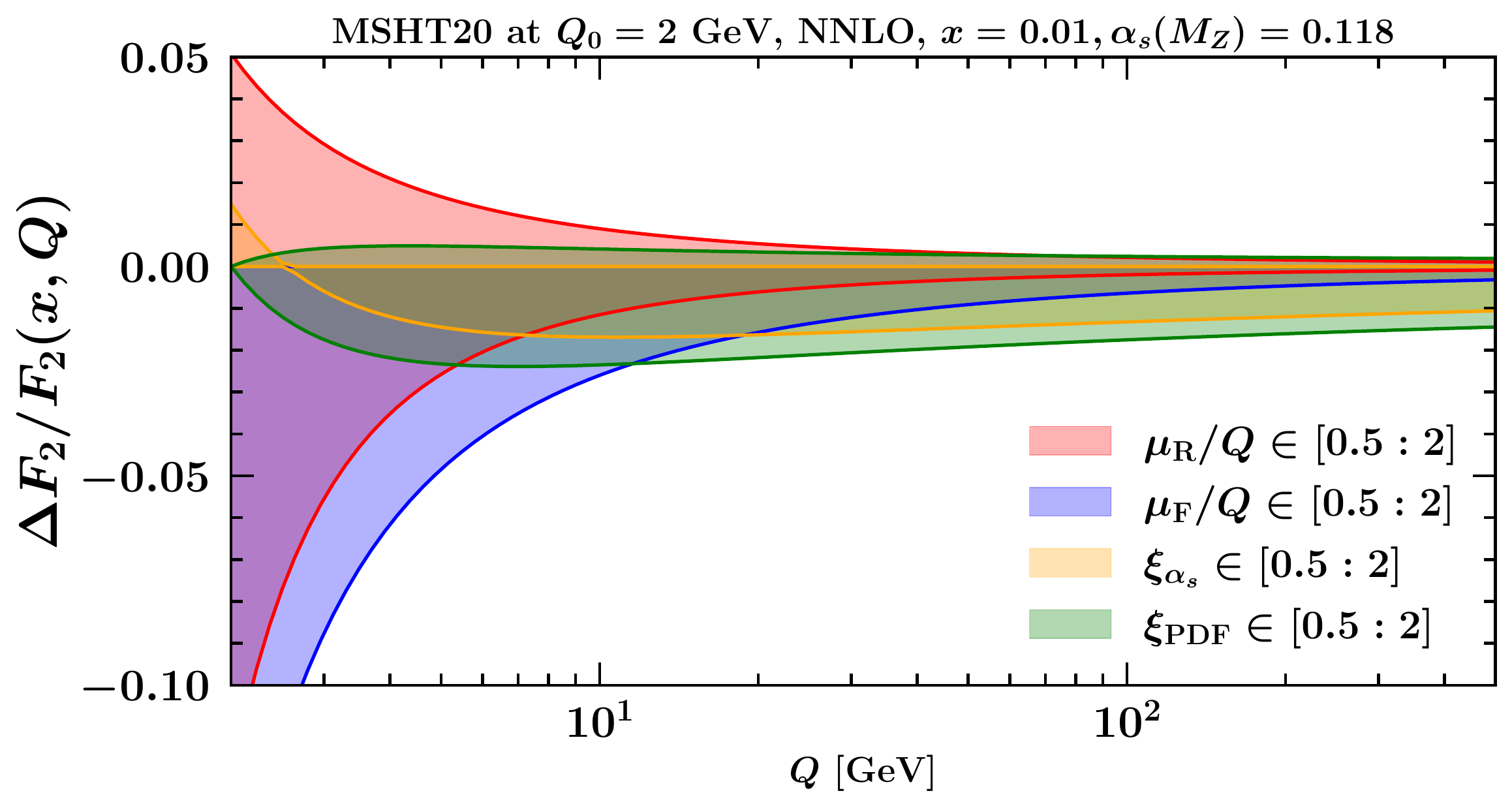}
\includegraphics[width=0.43\textwidth]{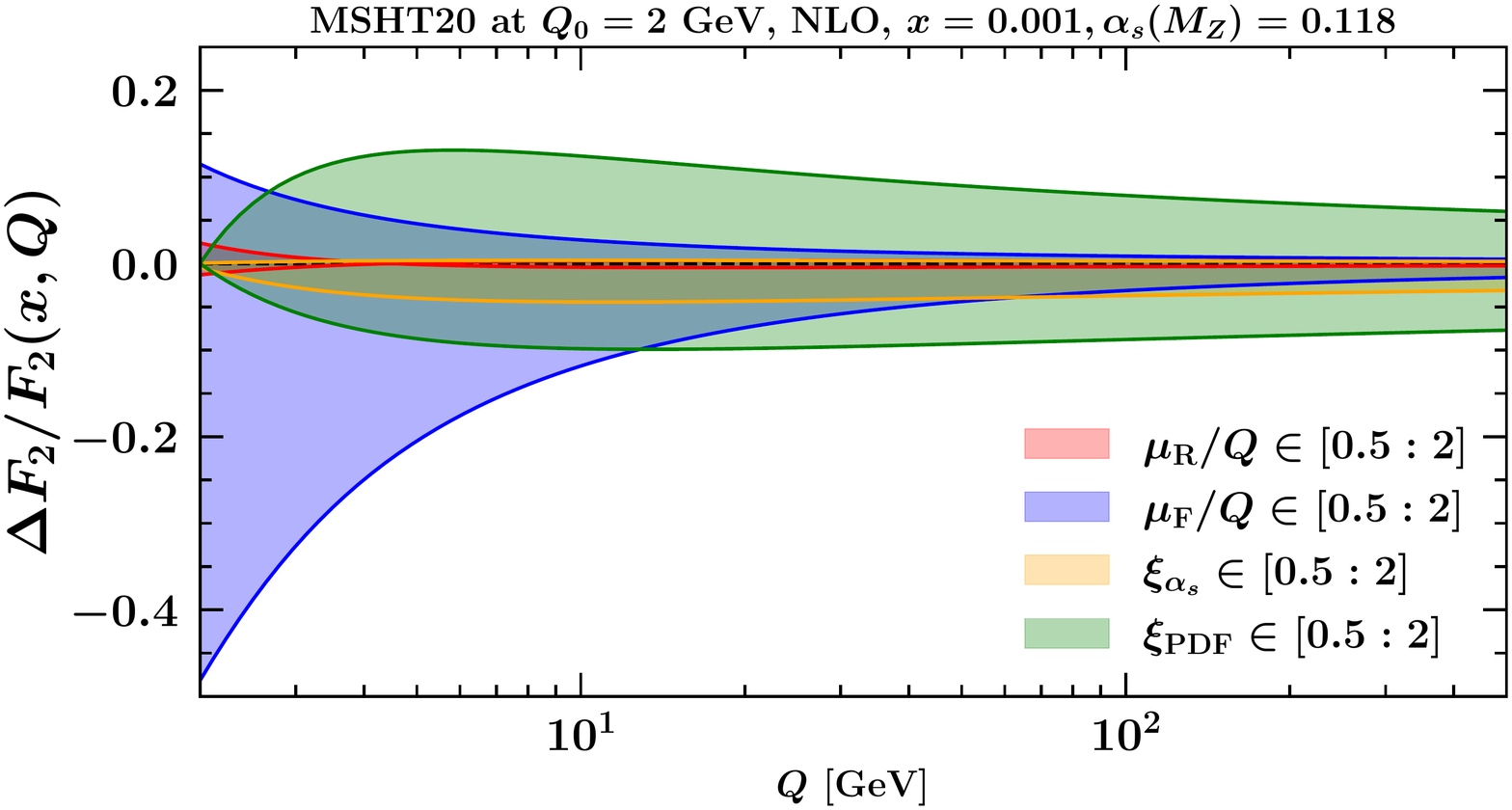} 
\includegraphics[width=0.43\textwidth]{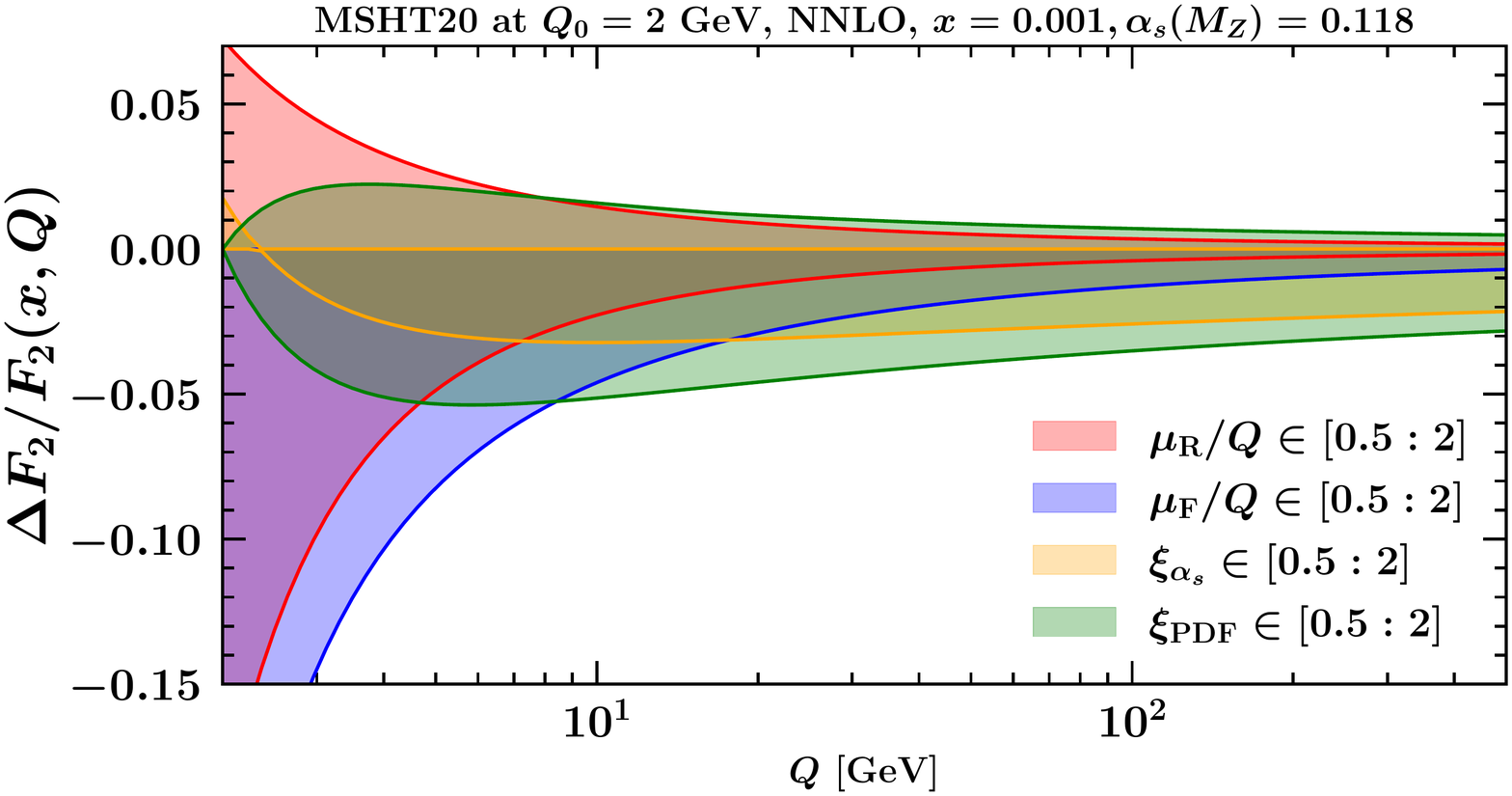}
\caption{$Q$-dependence of the relative variation 
$\Delta F_2 / F_2$ for $x=10^{-2}$ (top) and $x=10^{-3}$ (bottom), 
at NLO (left) and NNLO (right), associated with variations of renormalization and 
factorization scales, $\mu_R$ and $ \mu_F$, and resummation scale parameters $\xi_{\alpha_s}$ and $\xi_{\rm PDF}$.}
\label{fig:f2_muFmuR2resscales2}
\end{center}
\end{figure*}

The results shown in Fig.~\ref{fig:f2_muFmuR2resscales1} are for 
$Q = 10$ GeV. The importance of  resummation 
scale effects, 
 relative to that of renormalization and factorization scale effects, 
 increases with $Q$ (and  is 
  eventually becoming relevant for increasing $x$).  
  To illustrate the behavior of the uncertainty bands with varying $Q$, in 
  Fig.~\ref{fig:f2_muFmuR2resscales2} we plot the $Q$-dependence of the 
  relative variation  
  $\Delta F_2 / F_2$. The $\xi_{\rm PDF}$ contribution starts from 
  zero and grows rapidly with $Q$, remaining sizeable out to large $Q$ 
while the $\mu_F$ contribution is largest at low $Q$ and 
decreases with increasing $Q$. 
Analogously, the $\mu_R$ contribution is important at low $Q$ 
and decreases with $Q$, while the 
$\xi_{\alpha_s}$  is subdominant at low $Q$ but becomes relevant at high $Q$.

The  above numerical results 
indicate that RGE uncertainties are relevant in kinematical 
regions which influence current and forthcoming PDF extractions.  
They  
will also be relevant in regions explored at future 
lepton-hadron colliders~\cite{LHeC:2020van,Proceedings:2020eah}.  
As the $\xi_{\rm PDF}$ contribution remains significant at large $Q$, 
RGE uncertainties are  important for 
high-scale PDF probes such as very energetic jets and top quarks.  

To conclude, we note that 
the method we have described can be used     
 in processes 
sensitive to collinear as well as 
transverse momentum dependent (TMD)~\cite{Angeles-Martinez:2015sea}  
distributions.  
It will be applicable in extractions both of  PDF and of TMD.

\vskip 0.5 cm 

\noindent {\bf Acknowledgment.} We warmly thank N.~Armesto and the organizers of  the 
2022  Deep-Inelastic Scattering Workshop for putting up a very interesting 
conference and for allowing us to present these results from remote.

\end{document}